# Surface Structure of *In Situ* Cleaved Single Crystal $Bi_2Se_3$ Measured by Low Energy Ion Scattering

W. Zhou, H. Zhu, Jory A. Yarmoff*

*Department of Physics and Astronomy, University of California, Riverside CA 92521*

Bismuth Selenide is a two-dimensional topological insulator material composed of stacked quintuple layers (QL). The layers are held together by a weak van der Waals force that enables surface preparation by cleaving. Low energy ion scattering experiments (LEIS) show that $Bi_2Se_3$ cleaved under ultra-high vacuum (UHV) has a Se-terminated structure that is consistent with cleaving between QLs. Comparison of experimental data to molecular dynamics simulations confirms the Se-termination and provides an estimate of the surface relaxation.

___________

*Corresponding author, E-mail: yarmoff@ucr.edu

# I. Introduction

Topological insulators (TIs) are an emerging new class of materials that is of great interest due to the topological surface states (TSS) that make the surface conductive while the bulk is insulating [1]. The TSS are protected by time-reversal symmetry, which means that the surface carrier transport is not affected by non-magnetic impurities. The spins of the surface carriers are locked to their momentum protecting the carriers against backscattering since the spins have to be flipped over to make the carriers change the direction of their momentum. These unique properties make TIs very promising for spintronics [2,3], quantum computation [4] and other future technologies.

$Bi_2Se_3$ is the prototypical TI material, but its surface structure is still under debate. $Bi_2Se_3$ is a two-dimensional layered material, belonging to the class commonly known as van der Waals materials, with its basic building block being a quintuple layer (QL) ordered as Se-Bi-Se-Bi-Se [5], as illustrated in Fig. 1(a). The QLs are held together by a weak van der Waals force, so that $Bi_2Se_3$ is expected to naturally cleave between QL's to reveal a Se-terminated surface. There are, however, different terminations reported in the literature following cleaving, including Se-termination [6-9], Bi-termination [10,11], oxidized surface [11,12], coexisting Se- and Bi- terminations, either Se-termination or Bi-rich surfaces [6], and evolving surfaces [11].

In a recent study by our group [13], the atomic structure of $Bi_2Se_3$ surfaces prepared by *in situ* cleaving in UHV, *ex situ* cleaving in air and cycles of $Ar^+$ ion bombardment and annealing (IBA) were compared. It was shown that the *in situ* cleaved surfaces and those prepared by IBA are well ordered and Se-terminated, while samples cleaved in air have oxygen contamination and could be either Se-terminated or Bi-rich. In agreement with Ref. [10], it was hypothesized that the surfaces initially cleave to reveal a Se-termination, but that changes to the structure can occur afterwards. These changes are most likely caused by chemical reaction with atmospheric contaminants that preferentially adsorb at surface defect sites. The defects presumably result from the mechanical action of cleaving, and are most pronounced on smaller samples such as those utilized in Ref. [10]. Much larger samples were employed in Ref. [13] and in the present study, so that defect density is considerably reduced and the *in situ* cleaved samples are terminated with Se atoms.

A deeper understanding of the surface structure of these Se-terminated surfaces will help to better understand the origins of the TSS, as the atomic structure and electronic properties are highly dependent on each other [14-17]. The present paper utilizes low energy ion scattering (LEIS) to investigate the surface structure of $Bi_2Se_3$ samples prepared by *in situ* cleaving under pristine ultra-high vacuum (UHV) conditions. LEIS, which employs ions with incident kinetic energies between 0.5 keV and 10 keV, is a simple but powerful technique for surface structure analysis [18]. Due to shadow and blocking effects, ions that travel deeper than a few atomic layers are not able to escape from the surface. This makes LEIS especially surface sensitive and an ideal tool to determine the surface termination and atomic structure of TI materials.

## II. Experimental Procedure

Single crystals $Bi_2Se_{3.12}$ were grown using a slow cooling method [16] by melting mixtures of Bi shot (99.999%, Alfa Aesar) and Se shot (99.999+%, Alfa Aesar) in an evacuated quartz ampule (base pressure before being sealed $\approx 2\times10^{-6}$ Torr) with an inner diameter of 17 mm. The whole ampule was kept at 750°C for one day, slowly cooled to 500°C for 68 hours, and then annealed at 500°C for three days before being cooled to room temperature naturally.

The ingot was then cleaved with a razor blade (Fisher Scientific Company) to obtain flat samples that are a few mm in diameter.

The samples are mounted onto transferable holders so that new samples can be inserted and cleaved under vacuum to provide a fresh surface for each measurement. Each sample is attached to the holder with Ag paste (Epoxy Technology), and an Al bar is mounted on the top of the sample with a vacuum-compatible epoxy (Accu-Glass Products). The Ag paste and epoxy are cured by heating in a tube furnace at 100ºC for one-half hour. A load lock chamber with a sample transfer system (Thermionics) is attached to the main UHV chamber to enable quick introduction of new samples without the need for bakeout of the main chamber. After transferring the samples into the main chamber, they are cleaved *in situ* by knocking off the Al bar. After cleaving, the sample surfaces are flat and have a shiny appearance.

Low energy electron diffraction (LEED) and LEIS measurements are performed in the main UHV chamber, which has a base pressure of $2\times10^{-10}$ Torr. The rear-view LEED system (Princeton Research Instruments) is used to ascertain the sample order and sample orientation. Time-of-flight (TOF) LEIS spectra are collected using a pulsed $Na^+$ ion gun (Kimball Physics) and a micro-channel plate (MCP) detector, as previously described [19]. The ion gun points to the chamber center and is mounted on a turntable that can rotate about the vertical axis of the chamber. The sample is mounted horizontally at the end of a vertical x-y-z manipulator that allows it rotate azimuthally around the surface normal as well as around the chamber axis. The MCP detector is mounted at the end of a 0.57 m long flight tube, and collects scattered ions and neutrals with equal efficiency. There are two 3 mm-diameter apertures in the flight tube so that the acceptance angle in collecting TOF spectra is less than 1°. The TOF data collected in the present paper consist of the total scattered yield of ions and neutral atomic particles, so that neutralization effects do not need to be considered. During data collection, the ion beam, sample and detector are all in the same horizontal plane so that the incident polar angle, scattering angle and outgoing azimuthal angle can each be adjusted independently. The experiments reported here use 3 keV $Na^+$ ions with the incident beam directed along the normal to the sample surface. The ion beam is pulsed at 80 kHz by deflecting it across a 1 mm aperture, and a time interval counter (Stanford Research Systems) collects the flight times at which the projectiles arrive at the detector. TOF produces negligible beam damage because the small duty cycle of the pulsed beam leads to an effective incident ion beam current on the order of only pA, making it straightforward to keep the fluence well below 1% of a monolayer.

LEIS simulations are performed with Kalypso [20], a software package that uses molecular dynamics (MD) to simulate atomic collisions in solids. In the LEIS energy range, the incident ions are fast enough that only the repulsive interactions between projectiles and target atoms need to be taken into account while the interactions between target atoms can be ignored, which is known as the recoil interaction approximation (RIA) [18,21]. The repulsive Moliere-Firsov potential is used with a cutoff distance of 2.9 Å, and the Thomas-Fermi screening length is reduced by a correction factor, $c$, to best match the experimental data [22,23]. The $Bi_2Se_3$ target used for the simulations is a single QL with 5x5 atoms in each layer. The lateral lattice parameter and the unrelaxed first interlayer spacing are chosen as 4.14 Å and 1.59 Å, respectively, by averaging values from the literature [6,7,24-28] with the lowest and highest numbers for each parameter eliminated. The starting points of the top four surface interlayer spacings were set to the average of the two sets of structure parameters obtained via LEED and SXRD from ref. [7], but the spacing between the top two layers, which is the surface relaxation, was then adjusted to produce the best fit to the experimental data. The thermal vibrational amplitudes are calculated using a bulk Debye temperature of 200 K [29], while the mean square vibrational amplitudes of the top two atomic layers are enhanced isotropically by a factor of 1.6 by using a surface Debye temperature of 160 K.

**III. Results and Discussion**

After the sample is cleaved *in situ* under UHV, LEED patterns are used to check the quality of the crystal surface and to determine the azimuthal orientation. The 1x1 LEED pattern is a bright and sharp hexagon, indicating that the surface cleaved along the (001) plane and that it is well ordered [13]. The azimuthal orientation is defined here by setting $\varphi = 0°$ to be along the [210] direction, as illustrated in Fig. 1(b). The crystal structure itself is actually three-fold, however, despite the apparent six-fold symmetry of the LEED pattern. Thus, the LEED pattern alone is insufficient to distinguish the [210] direction ($\varphi = 0°$) from the [120] direction ($\varphi = 60°$). These directions can be distinguished, however, by comparing LEIS spectra collected along those azimuths, as explained below.

LEIS is a powerful technique for surface analysis. The incident energies are well in excess of surface bonding energies and the scattering cross sections are smaller than interatomic spacings, so the projectiles can be treated as though they undergo a series of isolated collisions with unbound target atoms located at the lattice sites, which is known as the binary collision approximation (BCA). If a projectile collides with only a single surface atom before backscattering from the sample, then the scattered projectile carries a kinetic energy that primarily depends on the projectile/target mass ratio and the scattering angle [18]. In this manner, each surface element that is directly visible to both the ion source and the detector produces a single scattering peak (SSP) in a LEIS spectrum.

The angular dependence of the SSP intensity is sensitive to the detailed atomic structure of the outermost few layers of a solid. The relationship of the angular yield to the structure is often understood using the concept of shadow and blocking cones [18,30]. A shadow cone is the region behind a surface atom that the projectile is excluded from, and it is calculated by mapping the trajectories of the incident ion beam around a single surface atom. A blocking cone is a similar concept, but applies to ions that have initially scattered from an atom below the outermost surface and is the angular region from which those projectiles are excluded from directly reaching the detector. The radii of the cones formed by low energy ions are typically several Å's, which is comparable to the interatomic distances. Thus, ions that travel deeper than a few atomic layers are unable to escape the surface, making LEIS very surface sensitive. As the scattered projectiles are deflected out of the shadow and blocking cones, the flux is increased at the cone edges. The details of the interaction of the cones with other atoms in the solid thus depend critically on the atomic structure, so that an analysis of the angular dependence of the scattered yield is an ideal tool for determining that structure.

A representative LEIS spectrum, collected from an *in situ* cleaved $Bi_2Se_{3,12}$ surface along the $\varphi = 60°$ azimuth using a scattering angle of $\theta = 115°$, is shown in Fig. 2(a). The flight time is converted to scattered energy using the known length of the flight tube and assuming that all of the scattered projectiles are $^{23}$Na. The peak at 1.1 keV is Se SSP and the one at 1.9 keV is the Bi SSP. This is consistent with the notion that projectiles scattered from heavier target atoms have a shorter flight time and thus a larger scattered energy. The Bi SSP is generally larger than the Se SSP because the scattering cross section is a strong function of mass. There is also a background of multiply scattered projectiles extending from approximately 700 eV to just above the Bi SSP. In actuality, the yield of multiply scattered ions continuously increases at smaller kinetic energies, but this is not seen in the spectrum because the MCP detector efficiency goes down rapidly as the energy of the detected projectiles falls below 1 keV [31]. Note that this decrease in MCP sensitivity also further reduces the intensity of the Se SSP relative to the Bi SSP.

A calculated energy spectrum is shown in Fig. 2(b) that was obtained with the same incident projectile energy and orientation used to collect the experimental data in Fig. 2(a), assuming a Se-terminated structure. The Se SSP in the calculated spectrum is at 1.2 keV and the Bi SSP is at 2.1 keV. The experimental SSP energies are approximately 100 to 200 eV below the simulated SSP energies due to continuous inelastic energy losses that are not included in the calculations [32]. The background in the simulated spectrum is very low, as compared to the experimental spectrum, because certain approximations are made to reduce the time needed to perform the calculations since only the SSP intensities need to be accurately reproduced. For example, a scattering event is terminated if the projectile spends too much time undergoing multiple collisions or if it penetrates too deeply below the surface. In addition, the target used is large enough to determine the SSP intensity, but is not large enough to account for many of the projectiles that undergo multiple scattering and still escape the sample. The small peak to the right of the Bi SSP is likely due to quasi-double scattering [33].

The measured azimuthal dependences of the Bi SSP intensity at scattering angles of $\theta$ = 115°, 125° and 135° are shown in Fig. 3 by the solid circles. The angular dependence of the Bi SSP is investigated, rather than that of the Se SSP, as the Se SSP does not contain information about the atomic structure below the outermost layer. Because the surfaces are terminated with Se, there is effectively no contribution to the Se SSP from Se located below the first layer due to shadowing and blocking. TOF spectra were collected at approximately every 5° from $\varphi$ = 0° to $\varphi$ = 120° for each of the three scattering angles. The SSP intensity at each angle was determined by subtracting the multiple scattering background and integrating the SSP. The background is approximated as a straight line connecting the edges of each peak, as illustrated by the dashed lines in Fig. 2(a). Figure 3 clearly confirms the mirror symmetry of the surface structure about the [120] azimuth ($\varphi = 60°$).

The simulated azimuthal dependence of the Bi SSP is shown as triangles in Fig. 3, along with the experimental data. As was done for the experimental measurement, the azimuthal dependence of Bi SSP intensity is obtained by increasing $\varphi$ in increments of 5º from 0° to 120°. The acceptance range for both the polar and azimuthal angles is set to 10°, as this produces much better statistics than does a 5° acceptance, while the results are almost unchanged. In addition, the three-fold symmetry and mirror symmetry around the crystal direction [$\bar{1}$20] ($\varphi$ = 60º) are all explicitly utilized to increase the counts at each angle.

The structure in the azimuthal dependence of the Bi SSP intensity is due primarily to projectiles that have scattered from second layer Bi atoms interacting with the blocking cones created by the outermost layer of Se. At normal incidence, only the top two layers of the sample can contribute to the Bi SSP intensity because the shadow cones, as illustrated in Fig. 1(a), prevent incident ions from directly impacting any other Bi atoms. Thus, only trajectories that originate by impacting the second layer need to be considered to interpret the data in Fig. 3. For example, if there is a Se surface atom positioned between the second layer Bi atom and the detector, then the blocking cones created by that Se atom will lead to a minimum in the scattered projectile yield along the direction of the Bi-Se bond. Because the projectiles scattered from second layer Bi are pushed towards the edges of the cones, the scattered yield will be enhanced along azimuthal directions that are between the surface Se atoms.

In this manner, consideration of the atomic structure can be used to qualitatively interpret the structure revealed by the azimuthal distributions. As the sample is rotated, as can be seen by inspection of Fig. 1(b), the outgoing direction for projectiles singly scattered from second layer Bi atoms moves through the three equivalent blocking cones of the nearest first layer Se atoms at $\varphi = 0°$, 120° and 240°. There are three other blocking cones equivalent to each other, which are due to the second nearest first layer Se atoms, which can interact with projectiles scattering from second layer Bi at $\varphi = 60°$, 180° and 300°. Because of the three-fold symmetry of the surface, the overall analysis can be limited to the region from $\varphi = 0°$ to 120° and only the two types of blocking cones at $\varphi = 0°$ and 60° need to be considered.

Figure 4 shows side views along the $\varphi = 0°$ and 60° azimuths, along with arrows that indicate the exit directions that correspond to the three scattering angles of $\theta = 115°$, 125° and 135°. The figure also displays blocking cones whose sizes were estimated by consideration of the measured and calculated azimuthal distributions, as discussed below. Along $\varphi = 0°$, and by symmetry along $\varphi = 120°$ or 240°, projectiles scattered from the second Bi layer are blocked at all three scattering angles by the first layer Se atoms, as these trajectories all pass close to the surface Se atoms leading to minima in the yield. This behavior is observed in Fig. 3 for all three scattering angles.

Applying this analysis to projectiles scattered at $\theta = 115°$, it is predicted that the outgoing beam will intersect the blocking cones at both $\varphi = 0$ and 60°, as these trajectories also pass very close to the surface Se atoms, resulting in minima at both of these angles, as observed. In addition, it is predicted that a broad maximum will occur around $\varphi = 30°$ as the projectiles that are scattered from $2^{nd}$ layer Bi atoms will be focused between the two blocking cones. This is precisely what is seen in Fig. 3(a) for both experiment and simulation.

For the projectiles scattered at $\theta = 125°$, the yield is intensified at $\varphi = 60°$. This can be explained by assuming that the outgoing trajectory is close to the edge of the blocking cone where the ion flux is enhanced. The blocking cone size in the schematic diagram in Fig. 4 was chosen to illustrate this. Thus, the Bi SSP intensity remains low at $\varphi = 0°$ but has maxima at $\varphi = 30°$ and $60°$, as observed in Fig. 3(b). Note that the difference in the intensities along the $\varphi = 0°$ and $60°$ azimuths at $\theta = 125°$ is how LEIS is used along with the LEED patterns to distinguish between the [210] and [120] azimuths when determining the absolute crystal orientation. Also, because the maximum at $\varphi = 30°$ results only from a single edge of the blocking cone at $\varphi = 0°$, it is sharper than the same maximum in Fig. 3(a) that results from a combination of two blocking cone edges.

For projectiles scattered at $\theta = 135°$, the outgoing projectiles are less affected at $\varphi = 60°$ as they are further away from the blocking cone. The outgoing direction still intersects the blocking cone at $\varphi = 0°$, however, causing a minimum in the Bi SSP intensity at $\varphi = 0°$ and maintaining the maximum at $\varphi = 30°$, as seen in Fig. 3(c). Since the outgoing trajectory is further from the edge of the blocking cone at $\varphi = 60°$, the Bi SSP intensity at $\varphi = 60°$ in Fig. 3(c) is less than it is for $\theta = 125°$ in Fig. 3(b).

Once the basic features of the azimuthal scans are understood, the parameters in the simulation are tuned to provide the best match to the experimental data. The screening length correction factor, $c$, which is commonly employed with the Moliere-Firsov potential, affects the size of shadow and blocking cones and therefore affects the peak positions in the azimuthal scans. Figure 5 shows simulations with different correction factors for normally incident 3 keV $Na^+$ scattered from $Bi_2Se_3$ at a scattering angle of 115°. The solid triangles show the calculated data while the solid circles are the same experimental data as in Fig. 3(a). Figure 5 shows that when $c$ increases, the maximum position of Bi SSP intensity at around 30° shifts to the right continuously while a second peak at around 45° shifts to the left continuously, and the factor $c$ = 0.8 gives the best match to the experimental data. This is because increasing $c$ increases the size of the blocking cones dramatically and therefore the azimuths corresponding to the peaks are further away from the atoms. A chi-square analysis is also used to compare each calculated distribution with the experimental data. Note that the data for $\varphi = 0°$ to 15° and the data for $\varphi = 55°$ to 60° were combined, as the counts are low. The results (not shown) indicate that $c = 0.8$ gives the lowest value of $\chi^2$, which agrees with the above analysis.

After determining the optimal value for $c$, the calculations are used to ascertain the surface relaxation, which is the spacing between the outermost and second atomic layers. Calculations were performed in which the interlayer relaxation between the top two atomic layers was adjusted from 6% (expansion) to -16% (contraction) relative to the bulk value of 1.59 Å. The calculated azimuthal dependence of the Bi SSP intensity at a scattering angle of 115° for various relaxations is shown in Fig. 6 along with the same experimental data that was shown in Fig. 3(a). When the relaxation changes, the ratio of the intensities of the overlapping peaks at 30° and 45° also changes. In addition, the SSP intensity at 60° increases when the contraction is too large. From the graph, it is obvious that the contraction should be larger than 1% because the peak at 45° will be too large when there is expansion. Conversely, the contraction should be smaller than 10% because the Bi SSP intensity at $\varphi = 60°$ increases obviously when there is too much contraction. Therefore, the relaxation is determined to be in the range of 1% to 10%, with a 2% contraction giving the best match to the experimental data.

This conclusion is consistent with first interlayer distances reported in literature, which are provided in Table 1. The reported values range from 1.51 Å to 1.62 Å. These relaxations, when given relative to the bulk value of 1.59 Å, range from a 2% expansion to a 5% contraction. The fact that this relaxation is very small is a reasonable conclusion, as the force between QL's is weak (van der Waals) so that the difference between a QL in the bulk and one at the surface should be minimal. Thus, the structure at the surface of a van der Waals material is expected to closely maintain the bulk structure.

## IV. Conclusions

Experimental ion scattering data shows that *in situ* cleaved $Bi_2Se_3$ is terminated with an intact QL. This is the expected result when cleaving between the van der Waals layers in the absence of contamination. This structure implies that Se is in the outermost atomic layer and Bi is in the second layer. Comparison of experimental and simulated LEIS data verifies this structure and concludes that there is only a slight relaxation between the outermost Se layer and the second layer Bi within the accuracy of a LEIS measurement.

## V. Acknowledgements

This material is based on work supported by, or in part by, the U.S. Army Research Laboratory and the U.S. Army Research Office under Grant No. 63852-PH-H.

**Table 1**. The spacing between the first and second atomic layer in single crystal Bi$_2$Se$_3$ as reported in the literature.

| Technique | First interlayer spacing (Å) | Relaxation | Reference |
|---|---|---|---|
| Low energy electron diffraction (LEED) | 1.56(3) | - 2% | dos Reis, et al. [7] |
| Surface x-ray diffraction (SXRD) | 1.51(5) | - 5% | dos Reis, et al. [7] |
| X-ray photoelectron diffraction (XPD) and holography (XPH) | 1.60(5) | + 1% | Kuznetsov, et al. [8] |
| Surface x-ray diffraction (SXRD) | 1.62 | + 2% | Roy, et al. [27] |
| Low energy ion scattering (LEIS) | 1.55 | - 2% | This study |

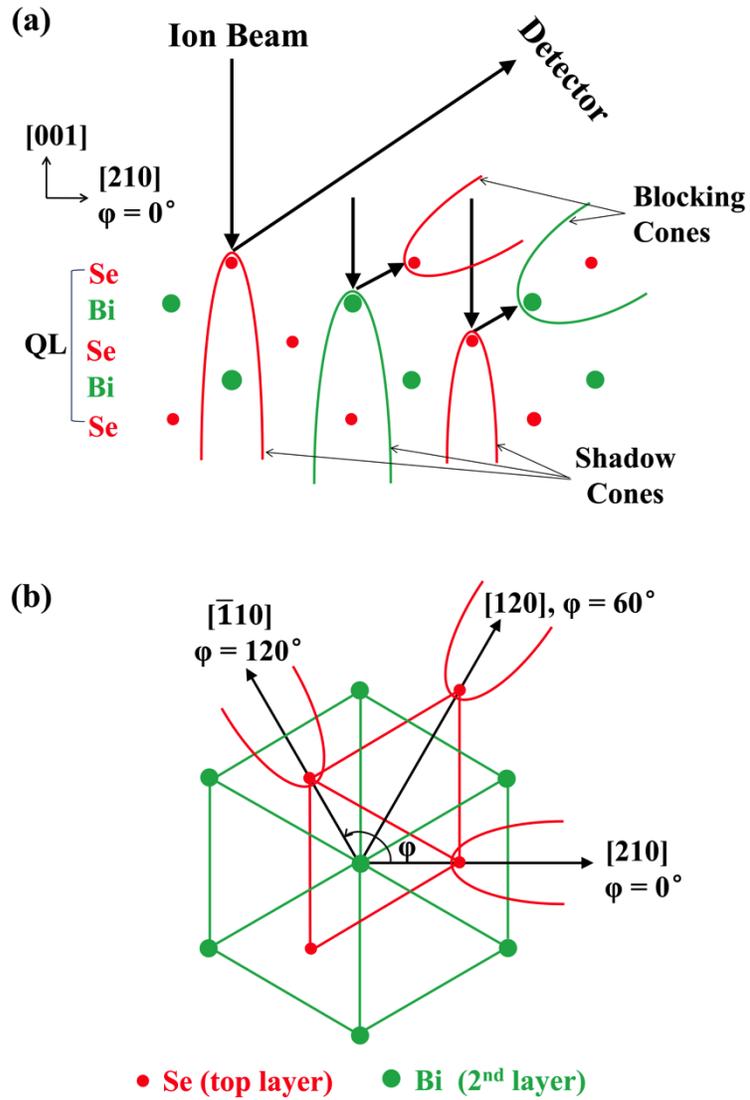

**Figure 1.** (a) A side view of the $Bi_2Se_3$ surface along the $(\bar{1}20)$ lattice plane, indicating the shadow and blocking cones formed by a normally incident ion beam with the detector placed at an angle of 35° from the surface plane. (b) A top view of the Se-terminated $Bi_2Se_3$ surface showing the outermost two atomic layers and indicating the azimuthal directions.

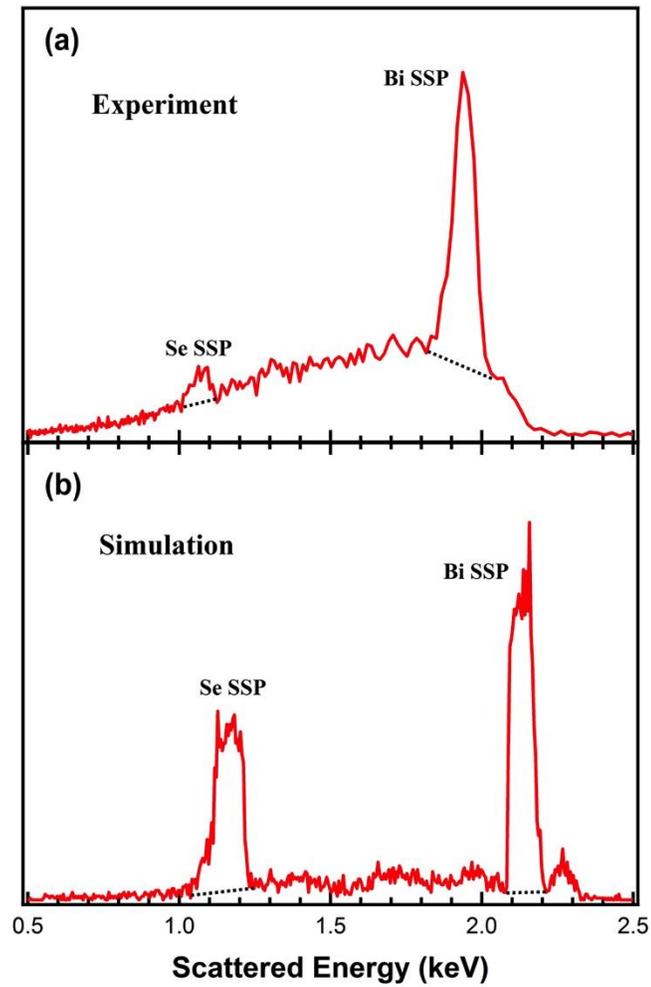

**Figure 2**. Typical measured (a) and calculated (b) energy spectra for normally incident 3.0 keV Na$^+$ scattered from Bi$_2$Se$_3$ at an angle of θ = 125°, collected along the φ = 60° azimuth.

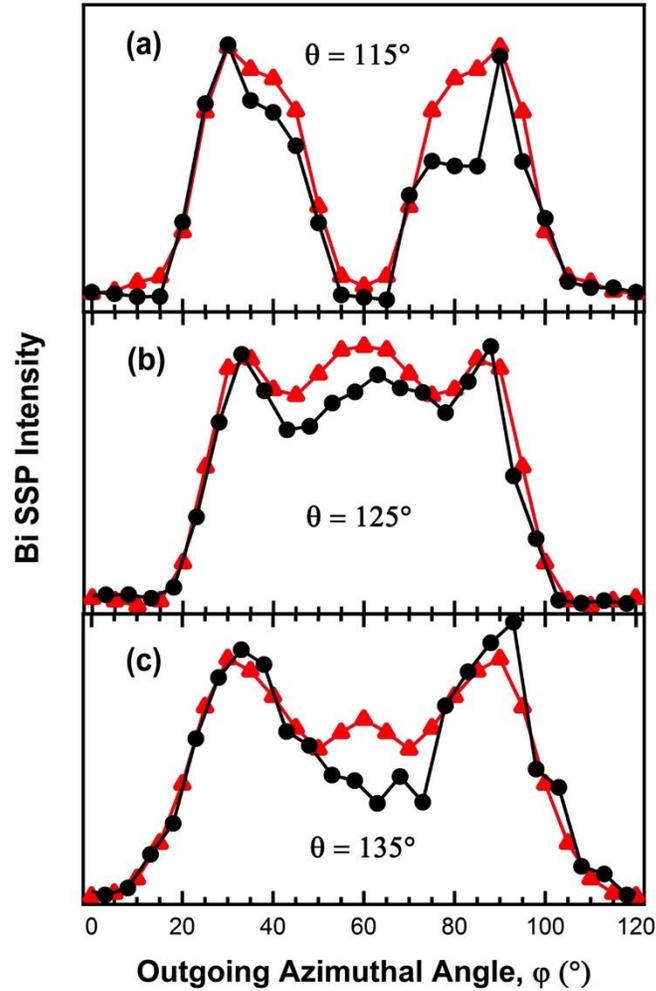

**Figure 3.** Experimental (circles) and simulated (triangles) azimuthal dependences of the Bi SSP intensity for normally incident 3.0 keV $Na^+$ scattered from $Bi_2Se_3$ at scattering angles of $\theta$ = 115°, 125°, and 135°. For the calculation, it is assumed that the first interlayer spacing is reduced by 3%.

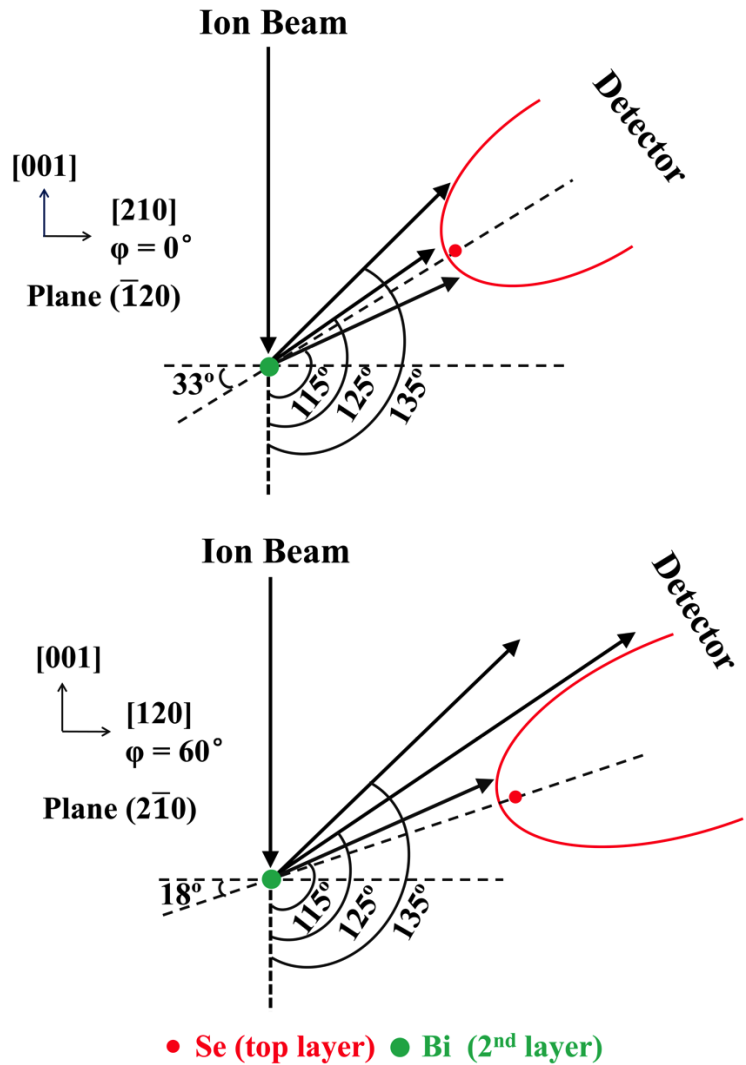

**Figure 4**. Side views of ion scattering trajectories for a normally incident beam that is scattered along the outgoing [210] and [120] azimuths. The arrows indicate three scattering angles of 115°, 125°, and 135°. The spacing between the first and second atomic layer is assumed to be relaxed by 2%.

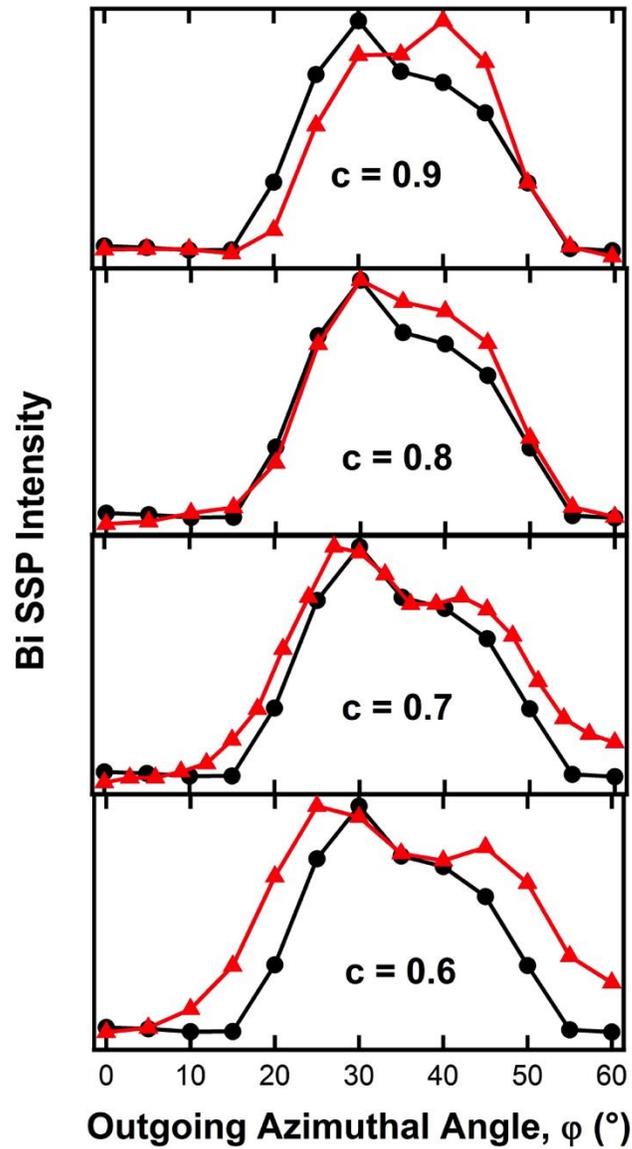

**Figure 5.** Calculated azimuthal dependence (triangles) of the Bi SSP intensity using the indicated screening length correction factor, *c*, in the Moliere-Firsov potential for normally incident 3.0 keV $Na^+$ scattered from $Bi_2Se_3$ at an angle of 115º. The first interlayer spacing was reduced by 3%. The solid circles show the experimental data.

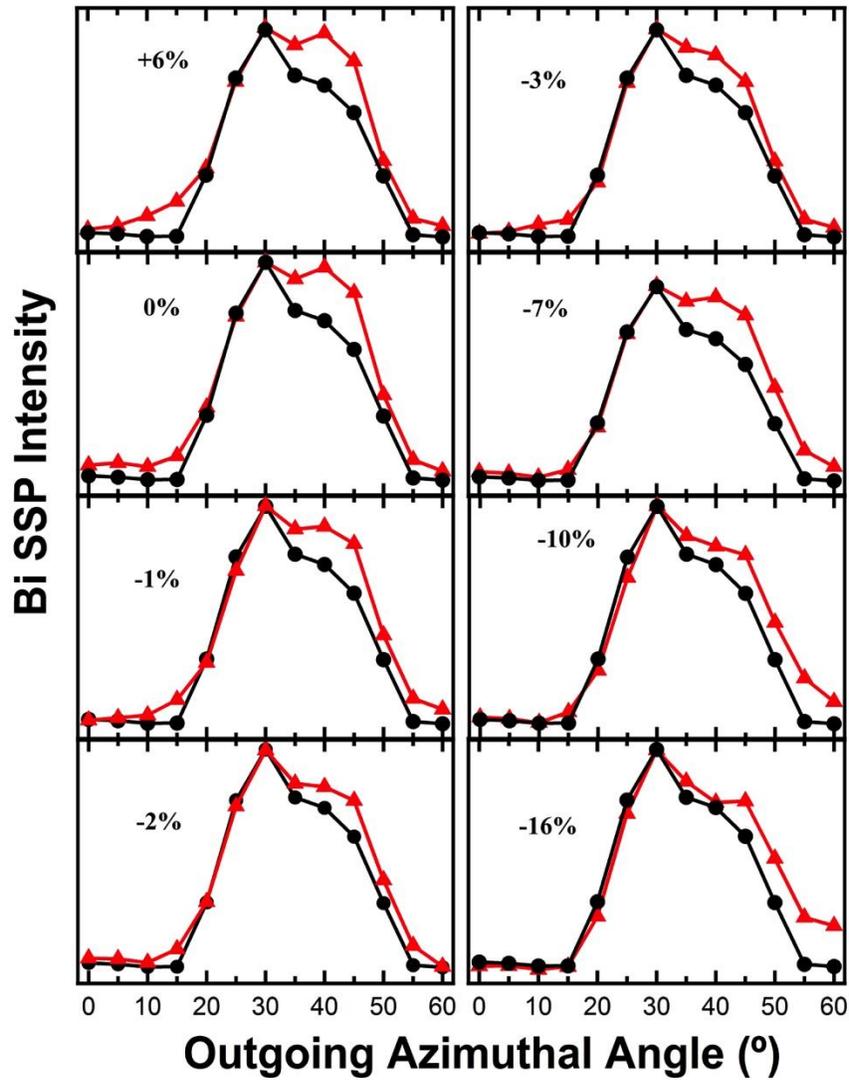

**Figure 6**. Calculated azimuthal dependences (triangles) of the Bi SSP intensity with the indicated relaxations between the top two atomic layers for normally incident 3 keV Na$^+$ scattered from Bi$_2$Se$_3$ at an angle of 115º. Negative and positive relaxation refers to contraction and expansion, respectively. The solid circles show the experimental data.